# Decorative coating or corrosion product?
# TEM and SEM Study on a Late Neolithic Axe to find origins of silver metal on the surface

**Khurram Saleem[1], Ulrich Schürmann[1], Lena Grandin[3], Christian Horn[3], Fritz Jürgens[2], Johannes Müller[2], Claus von Carnap-Bornheim[4],  Lorenz Kienle[1]**

[1] Institute of Materials Science and Engineering, CAU, Kiel, Germany

[2] Institute of Prehistoric and Protohistoric Archaeology, CAU, Keil, Germany

[3] Departement of Historical Studies, University of Gothenburg, Sweden

[4] Schleswig-Holstein State Museum, Schloss Gottorf, Schleswig, Germany

**Abstract**

Here we report on the analysis of a metallic Late Neolithic copper axe by means of Scanning Electron Microscopy (SEM) and Transmission Electron Microscopy (TEM). The axe was found at Eskilstorp, south-west Scania during archaeological excavations in autumn of 2015. It showed a hint of silver enrichment on the surface which is unusual for Late Neolithic axes from Pile type hoard. To identify the origin of the silver content, we extracted a thin lamella of the axe interior using Focused Ion Beam (FIB) technology to reach the internal structure to keep the process extremely minimally invasive. The results revealed the presence of porous external layer of copper oxide enriched with silver particles. It is shown with the corrosion test performed on copper-silver experimental replica that the silver enrichment is attributed to the selective dissolution of copper metal as a resut of corrosion process on the axe. The corrosion experiment is performed in the presence of an organic electrolyte such as Hemic Acid (HA) which replicates soil and natural water environment. In the scope of this research work, we found a strong evidence that the silver on the surface of the axe was not decrative layer and instead remained on the surface due to its more noble character while copper leached out into the soil due to corrosion.

## 1  INTRODUCTION

Archaeological artefacts provide a valuable insight into the human history and the determination of their original alloy composition is critical to understanding the evolution of materials over time. This information can aid in the development of theories about technological advancements and economic development, as well as provide insight into the social and cultural practices of the time period in which the artefact was fabricated. However, the analysis of archaeological artefacts present numerous challenges, particularly with regard to the effects of secondary alterations such as patination and corrosion. These alterations can cause surface modifications that obscure the underlying internal structure, making it difficult to accurately determine the original alloy composition of the artefact. One approach to overcoming these challenges is to conduct laboratory studies to investigate the factors that influence copper solubility and speciation in soil. These studies can include analyses of the effects of pH, degree of contamination, dissolved organic matter in soils, and soil type on the solubility of copper in soil. By understanding these factors, researchers can develop methods for more accurately determining the original alloy composition of archaeological artefacts. Additionally, the analysis of artefacts from different time periods can provide a comprehensive understanding of the evolution of materials over centuries. This approach can reveal important insights into the technological advancements and economic development of different societies and cultures [1][2]. Research into the original appearance of the ancient artefacts present major challenges owing to the corrosion or restoration of the artefacts [3]. In copper alloy objects, secondary alterations cause surface modifications due to patination and corrosion. As a result, differently colored layers appear on the surface of the artefact comprising of various chemical compounds. Black patina can form on any copper surface as a result of corrosion, however, intentional patina of the metal archaeological artefacts has also been found. Aucouturier et al. reported Intentional high temperature oxidation to obtain a dark patina on Egyptian and Roman artefacts [4].

Archaeological materials have been often investigated using numerous techniques such as X-ray Flourescence (XRF), Particle Induced X-ray Emission (PIXE), Scanning Electron Microscopy and X-ray Diffraction (XRD) [5]. With each technique having its advantages and limitations, a combination of these techniques is often utilized to attain detailed microchemical analysis. Corespondingly, the process of analysis is kept non-destructive since altering the structure of the valuable artefacts by analysis is not an option [6]. XRF, PIXE and SEM-EDS, are known for surface limited analysis techniques. In SEM, the chemical information from only a depth of 2.5 µm is attainable for copper based alloys even at 30 KeV [7]. Hulubei et al. reported on the classification of Apollonia and Dyrhachium silver drachmae (currency of Greek) found on Romanian territory by using SEM-EDS and PIXE. They acquired chemical composition of the drachmae and drew conclusions about the connections between the coins composition and historical aspects of the corresponding period. The authors associated hard economical and political situation with the increased Cu content of drachmae. [8]. Similarly, Ghoniem et al. utilized combination of SEM-EDS and XRD to reveal the corrosion products and contaminents such as Ca, C, S, Cl, Si, Al, Mg on Egyptian bronze statue.

The authors reported higher percentages of the actual composition such as Cu, Sn and Pb in the internal layer and lower concentration in the corrosion layer on the surface of the artefacts [9]. Therefore, to reach higher depths and attaining chemical information from inside the ancient artefacts, a more sophisticated sample preparation technique is crucial in archaeology. For this purpose Focused Ion Beam (FIB) milling could be utilized which is not frequently utilized on archaeological artefatcs so far. FIB offers the more advantageous means of obtaining access to the interna lcomposition with a non-noticeable invasion into the artefact [10].

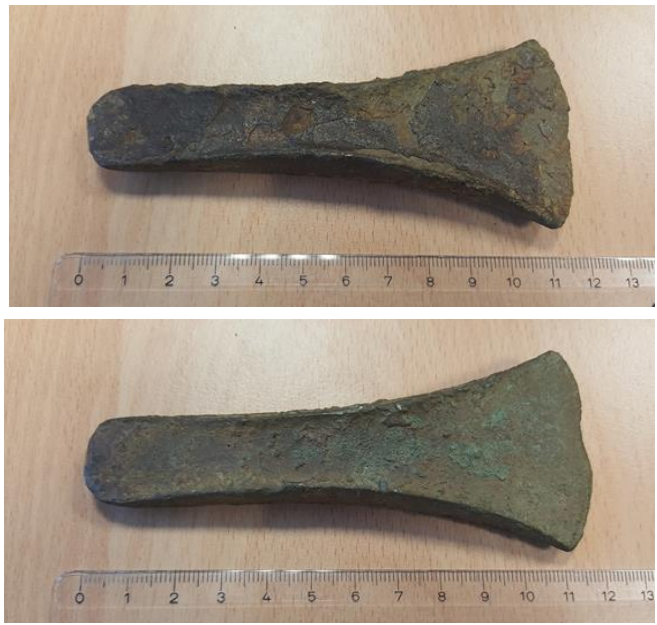

Figure 1. Neolithic copper axe with (a) front side
(b) backside

For nanoscopic elemental investigation Transmission Electron Microscopy (TEM) is performend on such thin slices of the material giving access to the internal structure of the artefacts. Vindel et al. performed TEM in combination with SEM-EDS to understand pre-collumbian copper metallurgy and specifically gilding and silvering methods. [11] [12].

In the scope of the research work presented in this paper, we show the analysis of the Late Neolithic copper axe from Pile type hoard. This work is aimed at developing a detailed understanding about the surface composition of the axe and identifying whether the silver enrichment on the surface is result of deliberate decorative treatment or it appeared as a result of corrosion phenomenon as the axe remains buried under the soil for thousands of years.

## 2 MATERIALS AND METHODS

### 2.1 *The Axe and sample preparaion*

The Neolithic Cu axe was found in 2015 at a medieval village Eskilstorp in South-West Scania in present day Sweden. According to the earlier observations in the field, the axe was found in a ploughed soil at the edge of a small drained bog [13]. Pile type axes were found in a hoard at Tygelsjö parish region located 5 km away from Eskilstorp in 2000 BC [13]. Arguably, one of the sides was laying facing downwards which might have caused difference in patination.

The axe is 115 mm high with a rounded butt width of 11 mm and convexed edge width of 48 mm. The thickness of the axe excluding the flanges is 14 mm. A few milimeter piece from the edge is chipped away from the cutting edde, interpreted as a possible combat damage [14]. Owing to the burial of the axe under the soil for nearly 4000 years, it is covered with soil residues and disfiguring crust of multicolored compounds. As a result, the surface layers comprising of secondary chemical alterations and contaminents are visible, not necessarily the original structure of the axe. The axe is covered by a greenish tint of oxidised copper alloy visible in Figure 1b whereas a darker contrasting surface is visible on majority part of the axe shown in Figure 1a and near the butt in Figure 1b. For TEM sample preparation, Focused Ion Beam (FIB) was used to extract two 12×12 µm sized slices from the surface of the axe. The extracted slices were welded onto a Nickel-Titnium (Ni-Ti) grid which was placed onto a TEM holder and loaded in the TEM coulmn for the analysis. For SEM analysis, a cross section cut from the edge on the butt of the axe was placed inside the SEM column.

Addtionally, to experimentally simulate the corrosion process, a copper/silver alloy was prepared (95 at. % of copper and 5 at. % of silver) and a corrosion experiment was conducted in an electrolyte which resembles the soil-like composition. The total weight of the alloy was about 5 g. To prepare the alloy, the silver wire was cut into smaller pieces, dipped in nitric acid (HNO3) for 5 minutes to clean the surface contaminations. Similarly, the copper beads were measured on a weighing scale and washed in nitric acid (HNO3) for 5 minutes. Both the metals were then packed in a glass ampoule. The ampoule was vacuum sealed using a heating stove. This ampoule was then placed in the furnace at 1200 °C for 5 hours. Then sample was left in the furnace to cool down overnight. The ampoule was then broken and the prepared alloy was retrieved resembling the shape of the ampoule. This alloy was then sectioned in two parts. Deliberate indentations were made on the sectioned alloy to mark the changes in the micro-structure of the same area after the corrosion experiment. To conduct the corrosion experiment, a solution of Humic acid with 56 mg/L in distilled water was prepared. The test solution was prepared in order to replicate conditions, similar to the soil environment. Humic acids (HA) are natural poly-electrolytes that occur in soil and natural waters because of the decay of plants and animal compounds, together with other biological species such as microorganisms in the environment[15]. Finally, the specimen was dried at room temperature, then degreased by acetone, rubbed with cotton wool, soaked in ethanol, dried at room temperature again and immediately immersed into the prepared solution. The prepared specimen remained immersed in the HA solution for 20 days at a room temperature. On the 20$^{th}$ Day, the sample was taken out of the test solution, weighted and left to dry at room temperature. The specimen was then mounted onto the SEM stage and analysed for the surface examination.

2.2 *Electron microscopy*

The axe was studied by means of EDS in a SEM Zeiss Gemini Ultra55 Plus. The secondary electron (SE) images, EDS elemental maps and point measurements were recorded from the regions of interest. TEM methods such as High Angle Annular Dark Field (HAADF) and Scanning TEM (STEM) were performed on the lamellae prepared via FIB. Electron Diffraction patterns (ED) were recorded to get information about the crystal structure and phases on the surface of the axe. TEM studies were performed in a FEI

Tecnai F30 G² STwin equipped with an EDS detector (Si/Li, EDAX). The TEM specimen should be less than 100 nm thin for precise investigation and as thin as possible for HRTEM [12]. Additionally the preparation should not affect microstructure mechanically or chemically and it must be representative of the true nature and morphology of the material.

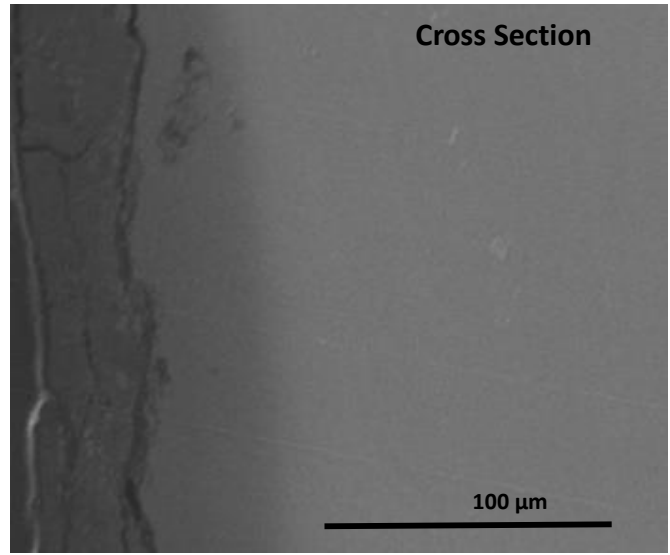

Figure 2. SEM on the cross section of the Neolithic Copper Axe showing the electron image

## 3 RESULTS AND DISCUSSION

*3.1 SEM investigations on the cross section from butt of the axe*

The results from the cross section of the axe shows a porous, deteriorated external rim on top of smooth homogenous core indicated with difference in contrast and morphology shown in Figure 2. The rim is estimated around 35-40 µm thick with varring thickness. Figure 3 shows the electron image and SEM-EDS elemental map conducted on the cross section of the axe. Elemental distribution shows the presence of Cu, Ag, O, and Cl in the Figure 3 (b—e). The steep surfaces and edges tend to be brighter than flat regions such as three-dimensional hill-like appearance shown in the Figure 3a. So, a thick layer is present on one side of the axe consisting of metallic compounds which is the outcome of the one-sided contact of the axe with the drained bog in which the axe was found. An interwoven pattern of Cu and particles of Ag on the surface are present. At a closer inspection it can be seen that the signals of Ag are rather isolated instead of homogenously distributed within the Cu and only present in the surface of the axe. So, a free copper layer is covered with isolated microscopic silver beads, sometimes present as spiral silver whiskers as shown in Figure 4, which Bella et al. regarded as typical silver corrosion texture [16].

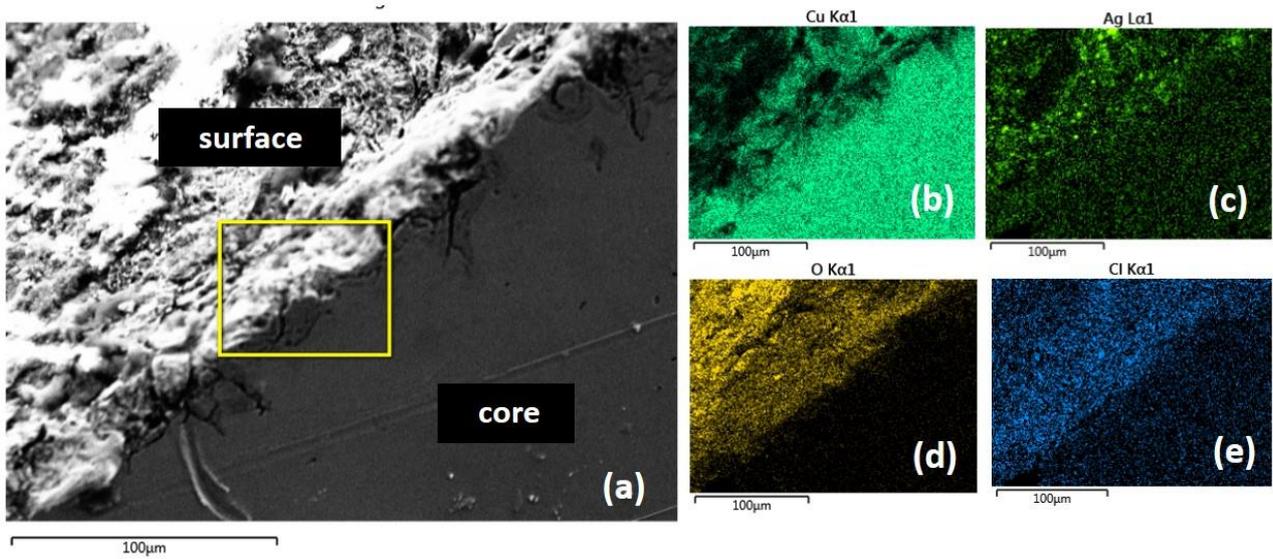

Figure 3. SEM image and elemental maps recorded at the cross section showing distribution of Cu, O, Cl and Ag

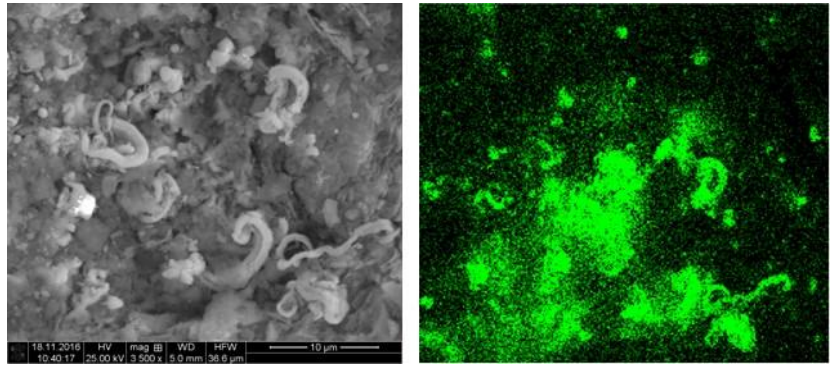

Figure 4. SEM electron image and elemental maps recorded at face A of the axe showing spiral silver structures

*3.2 TEM investigations on the FIB lamellae*

The TEM results from lamellae Aare shown in Figure 5 and from the lamella B are shown in Figure 6. The first figure shows the HAADF image, showing higher atomic number (Z) elements bright whereas the second figure shows bright field electron image, with higher Z elements shown as darker contrast. The alphabetical markings on the images represent the regions where STEM-EDS point measurements for elemental compositions were taken to attain the chemical information from the contrasting microstructural features. The elemental composition at those points are displayed in Table 1 (first and second lamellae), respectively. The dark field image in Figure 5 is significantly porous and the presence of Cu, Ag, Fe, P, O and Ca can be observed in the corresponding elemental map. It can also be seen that the silver micro particles are agglomerated in regions within the lamella and are surrounded by matrix of $Cu_2O$. This observation is confirmed by the electron diffraction pattern shown in Figure 7 taken from lamella B. The alphabetical markings on the HAADF image represent the regions where STEM-EDS point measurments for elemental compositions were taken. The elemental composition at those points is displayed in Table 1. The point measurements taken from two areas near the surface show the presence of heavy oxidation (ca. 56 at.%), P (ca. 6.6 at.%) and Ca (ca. 4 at. %). The bright contrasting region in the dark field image of the lamella A shows high weight percentage of Ag (ca. 53 at.%).

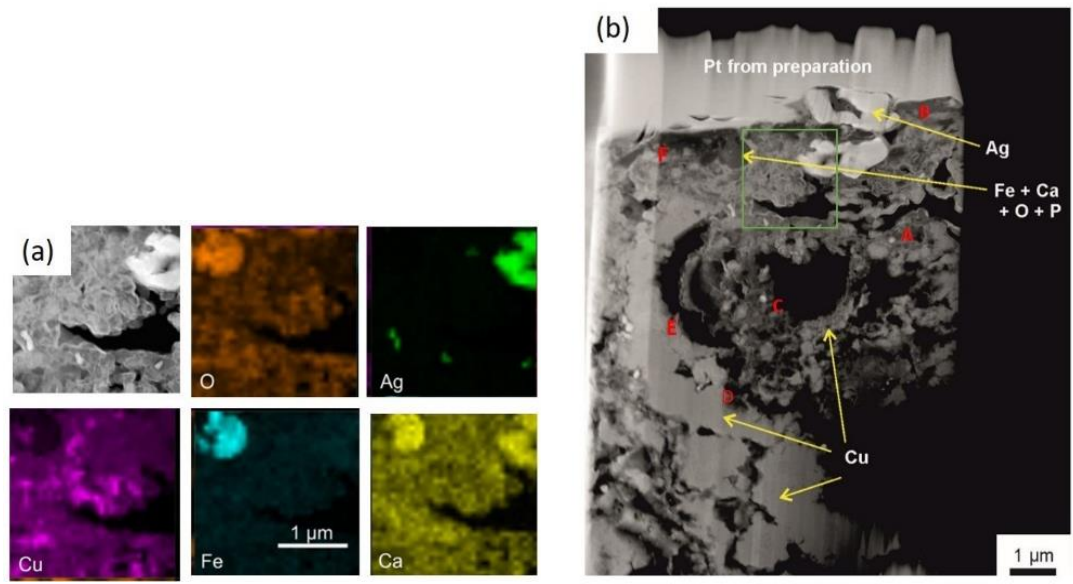

Figure 5. a) ) STEM- EDX elemental map of the region of interest and point measurements at marked regions b) HAADF of the FIB lamella from the copper Axe

Table 1 Representative result of STEM-EDS point measurements at different areas; the percentage uncertainity is less then 0.5% in all the measurements

| Area | Cu | Ag | C | O | Fe | P | Ca |
|---|---|---|---|---|---|---|---|
| **STEM-EDX Point Measurements from first lamella (at.%)** | | | | | | | |
| A | 11.3 | 52.3 | 36.2 | - | - | - | - |
| B | 5.9 | - | 18 | 56.6 | 1.8 | 6.6 | 4 |
| C | 41 | 44.3 | 7.2 | 2 | - | 5 | - |
| D | 97.1 | - | - | 2.8 | - | - | - |
| E | 96.3 | - | 1.9 | 1.7 | - | - | - |
| **STEM-EDX Point Measurments from second lamella (at.%)** | | | | | | | |
| A | 0.7 | 83.8 | 15.3 | - | - | - | - |
| B | 26.8 | 64.8 | 6.5 | 1.7 | - | - | - |
| C | 83 | - | 4.1 | 12.7 | - | - | - |
| D | 98 | - | - | 1.9 | - | - | - |

In the dark field image, contrasting features represent different elements and phases into the ca. 10 μm depth from the surface of the axe. The top layer is the protective platinum layer, part of the of FIB preparation process. Underneath the Pt layer, an elemental map from the actual surface of the axe reveals signals from Cu, Fe, Ca, P, O, C and Ag. The dark field image shows a porous structure with a similar distribution patterns of Cu, O and Ca. The FIB cut is very porous, clearly showing the deterioration of the original structure of the specimen due to secondary alterations and penetration of the contaminents possibly from the soil or enviroment. Similar to SEM-EDX, isolated beads of Ag distributed at different regions over the lamella can be seen. To attain more information, a second FIB cut was taken from a more intact region of the axe, shown in Figure 6 and corresponding STEM-EDS point measurements are shown in Table 1. The TEM image once again shows the contrast between different phases and structures in the lamella in Figure 6. It looks less porous compared to the lamella one, indicating less penetrated regions. That is also confirmed with eliminated signals from Fe and Ca and P in the point measurements represented for second lamella in Table 1. The diffraction patterns taken at relatively lighter contrast regions on the slice showed the presence of cuprite ($Cu_2O$) but also residuals of the original pure Cu. In the matrix of this cuprite, as indicated with yellow arrows, darker spots of Ag particles, owing to higher atomic number, are also evident. Representation of silver can be seen at two different lamellae taken from two different regions of the axe. The other contaminents such as Fe, Ca and P are limited to the surface and not found in the deeper (10-12 μm) regions. The distribution of Ag is not homogenous rather randomly present on the lamellae. The distribution of Cu and Cu oxide is further confirmed with the point measurement taken at areas B- E, displayed in Table 1.

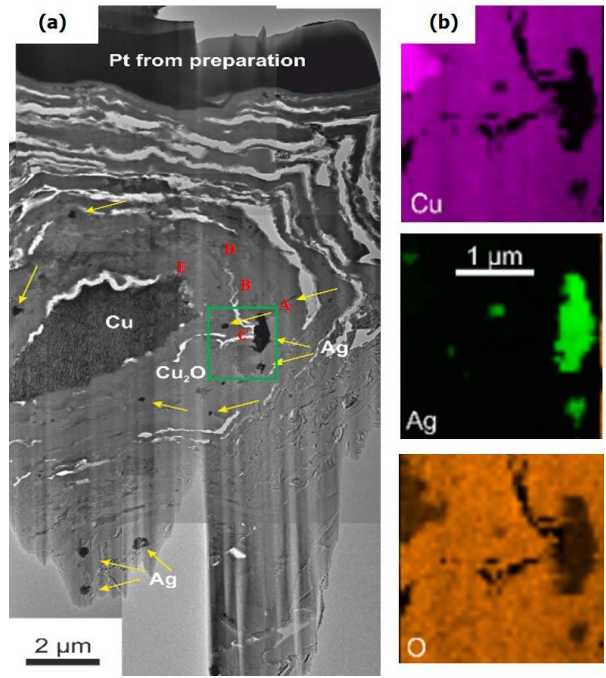

Figure 6. TEM image and STEM-EDS elemental maps of Cu, Ag and O.

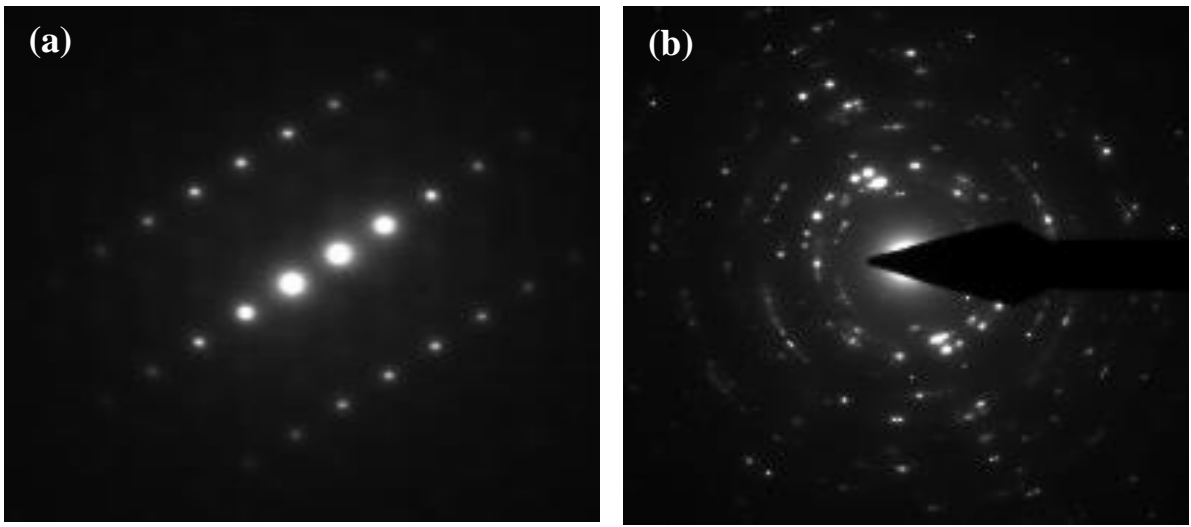

Figure 7. (a) Electron diffraction (ED) pattern of a sin crystalline Cu recorded at the darker region (b) polycrystalline ED pattern of the matrix $Cu_2O$

*3.3 SEM Analyses after the corrosion experiment*

Similar to the SEM results, the observations displayed in the TEM results show a cuprite rim on the surface of the axe with silver micro particles. Moreover, the SEM electron image showed typical silver whiskers (Figure 4) which grow as a corrosion product on the copper substrates. To experimentally simulate the corrosion process, a study was conducted in the scope of this research work in which the copper-silver alloy was prepared (94 at. % of copper and 6 at. % of silver) and a corrosion experiment was conducted in a Hamic Acid.

Figure 8 shows the EDS elemental mapping recorded on the prepared alloy (a) before the corrosion experiment and (b) after the corrosion experiment. The surface obtained after immersion in solution Hamic acid is relatively rough and it appeared to have been attacked on several regions by the corrosive solution. Further, it is demonstrated from the elemental mapping that a higher signals of silver are reported on the specimen after the corrosion experiment compared to pristine Cu-Ag specimen. The map shows certain regions of decreased copper content in the alloy after corrosion experiment when compared with pristine specimen. To quantify the elements present on the surface, EDX point measurements were recorded from 6 regions on the flat side of the specimen. The average values show increase in the silver content from 6 at. % in the pristine specimen surface to 13 at. % in the surface of the specimen after the corrosion test. The copper signals have decreased from 94 at. % to 83 at. %. Moreover, the oxygen content is increased from very low 0.4 at. % to 12.3 at. %. Hence, substantial changes in the surface morphology of the specimen have been observed after the corrosion experiment.

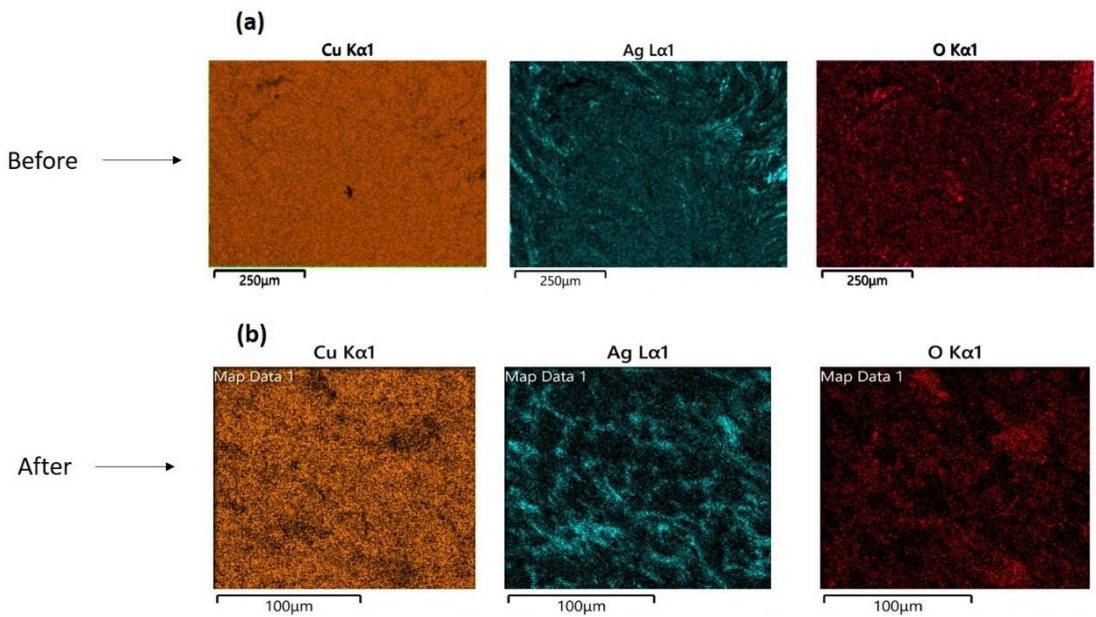

Figure 8. SEM EDX elemental mapping on the CuAg alloy (a) before the corrosion experiment (b) after the corrosion experiment

**Discussion**

From the SEM results of the axe, it is illustrated that the axe is compised of a copper core and the surface of the axe consists of an oxidised zone. This result is confirmed by the elemental mapping recorded at the cross section of the axe by the help of SEM EDS, shown in Figure 3. In the TEM results, it is demonstrated, that there is a presence of pure Ag microparticles over the whole FIB lamella. However, these silver particles are possibly less than 2%, when it is compared with the region where pure copper signal is recorded. These silver particles signal is more evident in certain isolated on the lamellae, shown in Figure 5 and 6. The structural information recorded in the form electron diffraction patterns shown in Figure 7, confirms the presence of $Cu_2O$ and Cu in the contrasting regions recorded on TEM image displayed in figure 6. Hence, the axe's surface is composed of a matrix of copper oxide which are surrounding Ag beads and crystals of pure copper. As it was documented by the initial metallographic analysis conducted by An et al. [17], fahlore copper ore was suggested to have been used in the production of this Eskilstorp axe. The fahlore type copper ore is reported to consist of small amounts (i.e. <1%) of impurities such as antimony, silver, arsenic and some time nickel.

Warraky et al. conducted the study on the corrosion of copper in chloride solutions. The authors demonstrated the formation of CuCl in the first step in high NaCl concentraions. In the second step the NaCl undergo hydrolysis and protective $Cu_2O$ forms on the surface. At the same time at some areas only a thin layer of $Cu_2O$ is present over trapped CuCl and at higher NaCl concentrations, the $CuCl_2^-$ starts to form, pit formation start at these regions and the formed chlorides ($CuCl_2^-$) can dissolve into the soil leaving copper depleted regions or pits. Further attacks on the protective oxide layer creates holes and pits which further attacks the underlying copper. As copper chloride is dissolved into the corrosive media, some cuprous chloride (CuCl) or cupric chloride ($CuCl_2$) stays onto the surface of the copper depending on the concentration of the chloride ions and pH of the environment. Once, the pits are formed, they continue to grow causing more copper to weaken and fall from the surface of the copper. Hence, the protective $Cu_2O$ formed on the surface of the axe eventually dissolved into the surrounding soil along with the formation cupric chloride CuCl. While copper is dissolved in the form of $CuCl_2^-$ in the pitting cracks, a proctective AgCl layer is formed on the surface which remains relatively stable. Ciacotich et al. reported on the selective dissolution of copper at localised regions in a electroplated copper-silver alloy [18]. Prajatelistia et al. conducted corrosion experiments at varying temperatures on 85% Cu, 15% Ag and 5% P alloy and reported on a similer phenomenon of selective dissolution of copper [19].

**Conclusion**

This study provides an intriguing insight in the surface, subsurface and the core structure of the axe. On the surface layer, the microscopic silver crystallites are distributed inside the oxidized copper matrix documented by EDS elemental maps and electron diffraction patterns. The copper from the surrounding regions of the silver was selectively dissolved where due to more noble nature of Ag it stayed inside the pits and holes on the surface. The Eskilstorp axe is heavily corroded with an extensive formation of cuprite. The silver impurity is isolated silver particles in the oxidised surface layer which were present as impurity in the source of the ore. There is enough evidence to state that no decorative layer was present on this Late Neolithic copper axe. Elements such as Ca, Al, Si, P, K and Fe can be associated with the soil remnants on the surface of the Eskilstorp axe and possibly correlated with the soil composition. However, soil from different regions differs greatly and it is difficult to predict the exact soil concentration. The correlation between the Ca and P in the STEM point measurements hints towards a possible presence of bone or clothing and its degradation products nearby at the site of the excavation.

.